\documentclass[aps,prl,twocolumn,groupeaddress,showpacs,showkeys]{revtex4-1}

\usepackage{epsfig}
\usepackage[usenames,dvipsnames]{color}

\newcommand{\tr}{{\rm tr}}
\newcommand{\cO}{{\cal O}}

\newcommand{\pslash}{{p \!\!\!/}}

\newcommand{\msbar}{{\overline{\rm MS}}}
\newcommand{\ripmom}{{\rm RI'\!/MOM}}
\newcommand{\smomgam}{{\rm RI/SMOM_{\gamma_\mu}}}
\newcommand{\GeV}{{\rm GeV}}

\newcommand{\eq}[1]{{(\ref{#1})}}

\begin{document}

\title{Precise $\overline{\rm MS}$ light-quark masses from lattice QCD
in the RI/SMOM scheme}

\author{Martin Gorbahn}
\affiliation{
  Technische Universit\"at M\"unchen,
  Institute for Advanced Study,
  Arcisstra{\ss}e 21,D-80333 M\"unchen, Germany}
\affiliation{
  Technische Universit\"at M\"unchen,
  Excellence Cluster ``Universe'', Boltzmannstra{\ss}e 2,
  D-85748 Garching, Germany}
\email{martin.gorbahn@ph.tum.de}
\author{Sebastian J\"ager}
\affiliation{
  University of Sussex, Department of Physics and Astronomy,
  Falmer, Brighton BN1 9QH, UK}
\email{S.Jaeger@sussex.ac.uk}

\date{April, 2010}

\begin{abstract}
We compute the conversion factors needed to obtain the
$\msbar$ and RGI up, down, and strange-quark masses
at next-to-next-to-leading order from the corresponding
parameters renormalized in the recently proposed 
RI/SMOM and $\smomgam$ renormalization schemes.
This is important for obtaining the $\msbar$ masses with the best
possible precision from numerical lattice-QCD simulations,
because the customary RI${}^{(}{}'{}^{)}$/MOM scheme
is afflicted with large irreducible uncertainties both on
the lattice and in perturbation theory.
We find that the smallness of the known one-loop matching
coefficients is accompanied by even smaller two-loop contributions.
From a study of residual scale dependences, we estimate the resulting
perturbative uncertainty on the light-quark masses to
be about 2\% in the RI/SMOM scheme and about 3\% in the
$\smomgam$ scheme. Our conversion factors are given in fully
analytic form, for general covariant gauge and renormalization
point. We provide expressions for the associated anomalous dimensions.

\end{abstract}


\maketitle

Lattice QCD has, in recent years, seen important progress on several fronts:
there exist lattice regularizations preserving exact chiral symmetry
in the limit of vanishing quark masses, while algorithmic and
technological advances have put lattices fine enough to simulate
physical light-quark masses within reach. As a result, nonperturbative
results in the physics of light quarks with a precision of a few percent
or better become achievable with current or upcoming simulations
\cite{latticereviews}.
These include the masses of the light quarks, as well as hadronic matrix
elements such as $B_K$, figuring prominently in the
unitarity-triangle analysis.
At such high precision, choices of renormalization scheme and
associated perturbative higher-order effects become
an important source of uncertainty. Two standard methods have
emerged: the use of momentum-space subtraction schemes that can be
nonperturbatively implemented on a lattice \cite{Martinelli:1994ty}
and the Schr\"odinger-functional method \cite{sf}, where so-called
renormalization-group-invariant (RGI) masses and matrix elements are
obtained via a direct implementation of the renormalization
group on the lattice. Within the former approach, parameters need
a further conversion to purely perturbative schemes such as
$\msbar$ \cite{Bardeen:1978yd}, where short-distance QCD and new-physics effects
are best tractable.

It has recently
been realized that the standard  RI${}^{(}{}'{}^{)}$/MOM prescription 
suffers from a strong sensitivity to IR effects \cite{Aoki:2007xm}, which
has become the dominant source of uncertainty on the lattice. This is paralleled
by unusually large higher-order terms in the perturbative conversion
factors \cite{Chetyrkin:1999pq}.
A modified scheme with much better IR behaviour has been
recently proposed and called RI/SMOM \cite{Sturm:2009kb}.
In this work, we study the renormalization of the pseudoscalar
(non-singlet) density, which by virtue of chiral symmetries is related
to the renormalization of the quark mass, and
obtain the next-to-next-to-leading-order (NNLO, two-loop)
conversion factor allowing to obtain
$\msbar$ light quark masses from their counterparts renormalized
in the RI/SMOM scheme, or its variant $\smomgam$,
as `measured' on the lattice.
We find much smaller perturbative corrections than
in the RI${}^{(}{}'{}^{)}$/MOM case, extending one-loop findings
in \cite{Sturm:2009kb} and implying percent-level uncertainties on
the $\msbar$ masses.

\section{RI$'$/MOM, RI/SMOM, and {\boldmath $\smomgam$}}
In the RI$'$/MOM renormalization scheme for the quark field
and mass, two conditions \cite{Martinelli:1994ty}
\begin{eqnarray} \label{eq:ripwf}
  \lim_{m_R \to 0} \frac{1}{12\,p^2} \tr [S_R^{-1}(p) \pslash]
  \Big|_{p^2=-\mu^2} &=& - 1 ,
\\  \label{eq:ripmass}
  \lim_{m_R \to 0} \frac{1}{12\,m_R} \tr [S_R^{-1}(p)]
  \Big|_{p^2=-\mu^2} &=& 1 ,
\end{eqnarray}
are imposed on the inverse quark propagator $S_R^{-1} = Z_q^{-1} S_B^{-1}$.
The bare quark propagator $S_B$ is defined through
(our notation closely follows \cite{Sturm:2009kb})
\vspace*{-1mm}
\begin{equation}
  -i\,S_B(p) = \int \!{\rm d}^4 x\,e^{i p x}
     \langle T(\psi_B(x) \bar \psi_B(0)) \rangle ,
\end{equation}
and the traces are over colour and Dirac indices.
\eq{eq:ripwf} and \eq{eq:ripmass} determine the
renormalization constants $Z_q$ and $Z_m$ relating bare and
renormalized field and mass,
$\psi_R = Z_q^{1/2} \psi_B$ and $m_R = Z_m m_B$ .
Both renormalization constants depend implicitly on the regulator
(lattice, dimensional regularization, etc.) and on the gauge coupling
and the gauge parameter.
A virtue of the RI$'$/MOM scheme is that it can be implemented
nonperturbatively on the lattice as well as in dimensionally
regularized continuum perturbation theory. 
The RI$'$/MOM field and mass can then be
converted perturbatively to the $\msbar$ scheme
via $\psi_R^\msbar = (Z_q^\msbar/Z_q^{\ripmom})^{1/2} \psi_R^{\ripmom} $
and $m_R^\msbar = Z_m^\msbar/Z_m^{\ripmom} m_R^{\ripmom} $ ,
where all renormalization constants have to be computed with the same
(but otherwise arbitrary) regulator.
Both conversion factors are known to three-loop accuracy
\cite{Chetyrkin:1999pq,Gracey:2003yr}.
However, the perturbation series does not converge well, and this
constitutes a drawback of using the RI$'$/MOM scheme for extracting
light-quark masses from lattice simulations. Another issue is the
influence of non-perturbative long-distance physics. This is most
clearly seen by considering (non-singlet) axial-current Ward
identities such as
\begin{eqnarray}
  \lefteqn{
q_\mu \Lambda^\mu_{A,B}(p, p') =
} \nonumber   \label{eq:WIax}
\\
&&
          S_B^{-1}(p')\gamma_5 + \gamma_5 S_B^{-1}(p)
          + i (m_{u,B}\! + \! m_{s,B}) \Lambda_{P,B}(p, p') , \quad \;
\end{eqnarray}
where $q\equiv p-p'$,
and the bare vertex functions
$\Lambda^\mu_{A,B}$ for the axial current and
 $\Lambda_{P,B}$ for the pseudoscalar density
are defined through
\begin{eqnarray}
\lefteqn{
  S_B(p') \Lambda^\mu_{A,B}(p,p') S_B(p) }
 \\ &=& \! \!  \int \!\!{\rm d}^4 x\,{\rm d}^4 y\,e^{i p' x} e^{-i p y} \,
  \langle T([i \bar u_B \gamma^\mu \gamma_5 s_B](0)\, u_B(x)\, \bar s_B(y))
  \rangle , \; \nonumber
\\[2mm]
\lefteqn{
  S_B(p') \Lambda_{P,B}(p,p') S_B(p) }
  \nonumber \\
     &=& \! \!  \int \!\!{\rm d}^4 x\,{\rm d}^4 y\,e^{i p' x} e^{-i p y} \,
  \langle T([i \bar u_B \gamma_5 s_B](0)\, u_B(x)\, \bar s_B(y))
  \rangle . \;\;
\end{eqnarray}
\eq{eq:WIax} holds for a regulator which respects chiral
symmetry (in the limit $m_B \to 0$). This is the case for certain
lattice regularizations and for dimensional regularization with anticommuting
$\gamma_5$.
(The use of anticommuting $\gamma_5$ is unproblematic
here as \eq{eq:WIax} and the formulae below
do not involve closed traces containing odd powers of $\gamma_5$.)
To preserve \eq{eq:WIax} under renormalization, the axial current must
not be renormalized, and the renormalization
constant $Z_P$ of the pseudoscalar density must satisfy
$  Z_P = Z_m^{-1} $,
where $Z_P$ can be fixed by imposing the
condition
\begin{eqnarray}  \label{eq:ZPSMOM}
\lambda_R(p^2, {p'}^2, q^2)
  &=& Z_q^{-1} Z_P\, \lambda_B(p^2, {p'}^2,q^2)
\nonumber \\ 
 &\equiv& Z_q^{-1} Z_P\, \tr[\Lambda_{P,B}(p, p') \gamma_5]
  \stackrel{!}{=} 12  \qquad
\end{eqnarray}
at a suitable subtraction point. The choice  $p^2={p'}^2=-\mu^2$, $q^2=0$
corresponds to \eq{eq:ripmass}. But at $q^2=0$, $\Lambda_{P,B}(p, p')$
receives contributions from the kaon (pseudo-Goldstone)
pole, which diverge in the
chiral limit $m_R \to 0$ \cite{Martinelli:1994ty},
and is sensitive to condensate effects
suppressed only by
$(\Lambda_{\rm QCD}/\mu)^2$ \cite{Aoki:2007xm}.
In \cite{Sturm:2009kb}, a modified renormalization scheme,
termed RI/SMOM, was proposed, which is less sensitive to these effects.
In that scheme, \eq{eq:ZPSMOM} is imposed at the symmetric
point $p^2={p'}^2=q^2=-\mu^2$.
Following \cite{Sturm:2009kb}, we will consider a more general
kinematic configuration
$p^2={p'}^2=-\mu^2$, $q^2= - \omega \mu^2$ below,
and define conversion factors
\begin{equation}
C_q^{\rm RI/SMOM} = C_q^{\rm RI'/MOM} \! = \! \frac{Z_q^\msbar}{Z_q^{\rm RI'/MOM}}
  \!=\! \frac{12\,\mu^2 Z_q^\msbar}{\sigma_{B}(-\mu^2)} ,
\label{eq:CqSMOM}
\end{equation}
\begin{equation}
C_m^{\rm RI/SMOM}(\omega) \!=\! \frac{Z_m^\msbar}{Z_m^{\rm RI/SMOM}(\omega)}
  \! = \! \frac{Z_m^\msbar \sigma_{B}(-\mu^2)}
            {\mu^2\,\lambda_{B}(\!-\mu^2,\!-\mu^2,\!-\omega \mu^2\!)} ,
       \label{eq:ZmSMOM} 
\end{equation}
where
$     \sigma_{B}(p^2) \equiv \tr[S_B^{-1}(p) \pslash] $.
The right-most expression in  \eq{eq:ZmSMOM} has a
straightforward perturbation expansion. Moreover,
in \cite{Sturm:2009kb} a variant scheme $\smomgam$
was introduced where
the field-renormalization condition \eq{eq:ripwf} is replaced by
the requirement
\begin{eqnarray}  \label{eq:ZqSMOMgam}
\tilde \lambda_R(p^2, {p'}^2, q^2)
  &=& Z_q^{-1}\, \tilde \lambda_B(p^2,{p'}^2,q^2)
\nonumber \\ 
 &\equiv& Z_q^{-1}\, \tr[\Lambda_{A,B}^\mu(p, p') \gamma_5 \gamma_\mu]
  \stackrel{!}{=} 48 , \qquad
\end{eqnarray}
which implies conversion factors
\begin{eqnarray}
  C_q^{\smomgam}(\omega) &=& 
     \frac{48\, Z_q^\msbar}{\tilde \lambda_{B}(-\mu^2,-\mu^2,-\omega \mu^2)} ,
                                            \qquad \label{eq:CqSMOMgam}
\\
  C_m^{\smomgam}(\omega, \omega') 
     &=& \frac{Z_m^\msbar \,\tilde \lambda_{B}(-\mu^2,-\mu^2,-\omega' \mu^2)}
            {4\,\lambda_{B}(-\mu^2,-\mu^2,-\omega \mu^2)} . \quad
       \label{eq:CmSMOMgam}
\end{eqnarray}
The schemes for field and mass are converted as
\begin{equation}
   \psi^{\msbar} = \left(C_q^X\right)^{1/2} \psi^X ,
\quad 
   m^{\msbar} = C_m^X\, m^X ,
\end{equation}
where $X=$ RI/SMOM or $\smomgam$.

We note that $C_q^X$ and $C_m^X$ depend
on $\ln \mu^2/\nu^2 \equiv \ln r$, where $\nu$
is the dimensional renormalization scale, and implicitly
on $\nu$ through the scale dependence of $\alpha_s$ and the
gauge parameter $\xi$.
Setting $\mu \equiv \nu$
allows relating the anomalous dimensions in the RI/SMOM schemes to
those in the $\msbar$ scheme
\cite{Tarrach:1980up,Tarasov:1982gk,Chetyrkin:1997dh,Vermaseren:1997fq}
according to
\begin{eqnarray}
   \gamma_m^X = \gamma_m^\msbar    \label{eq:admm}
         - \Big[ \frac{\beta(\alpha_s)}{4}
            \frac{\partial}{\partial \left(\frac{\alpha_s}{4\pi}\right)}
          + \delta(\alpha_s,\xi) \frac{\partial}{\partial \xi}\Big]
            \ln C_m^X \Big|_{r=1} \!, \;\;\;
\quad
            \\
   \gamma_q^X = \gamma_q^\msbar    \label{eq:admq}
         - \Big[ \frac{\beta(\alpha_s)}{4}
            \frac{\partial}{\partial \left(\frac{\alpha_s}{4\pi}\right)}
        + \delta(\alpha_s,\xi) \frac{\partial}{\partial \xi} \Big]
           \ln C_q^X \Big|_{r=1} \!. \;\;\;
\quad
\end{eqnarray}
Here we use the definitions (which conform to \cite{Sturm:2009kb})
\begin{eqnarray} \label{eq:admdef}
  \gamma_m^Y m^Y &=& \mu^2 \frac{{\rm d}}{{\rm d}\mu^2} m^Y, 
\qquad
  \gamma_q^Y \psi^Y = 2 \mu^2 \frac{{\rm d}}{{\rm d}\mu^2} \psi^Y , \\
  \beta(\alpha_s) &=& \mu^2 \frac{{\rm d}}{{\rm d}\mu^2}
  \frac{\alpha_s}{\pi} ,
\qquad
  \delta(\alpha_s,\xi) = \mu^2 \frac{{\rm d}}{{\rm d}\mu^2} \xi ,
\label{eq:betadef}
\end{eqnarray}
with $Y=\msbar$ or RI/SMOM or $\smomgam$.

\section{NNLO computation}
We now compute the conversion factors
to $\cO(\alpha_s^2)$
in dimensional regularization ($d=4-2\epsilon$).
Let us denote
\begin{eqnarray}
  \sigma &=& - 4 N_c p^2 + \sigma^{(1)} + \sigma^{(2)} + {\cO}(\alpha_s^3) ,
 \\
  \lambda &=& 4 N_c + \lambda^{(1)} + \lambda^{(2)} + \cO(\alpha_s^3) ,
 \\
  \tilde \lambda &=& 4 \, d \,  N_c + \tilde \lambda^{(1)}
      + \tilde \lambda^{(2)} + \cO(\alpha_s^3) ,
\end{eqnarray}
where the superscripts denote the loop order.
$\sigma^{(1)}$, $\lambda^{(1)}$, and $\tilde \lambda^{(1)}$
have been evaluated in \cite{Sturm:2009kb}. For the present
computation, we also need their $\cO(\epsilon)$ parts, which will affect
the $\cO(\alpha_s^2)$ results for $C_q$ and $C_m$. Taking the traces
and employing partial fractions, we obtain
\begin{eqnarray}
  \lefteqn{\sigma_{B}^{(1)}(p^2) =
    4 N_c C_F \, (-p^2)^{1-\epsilon}\,\frac{\alpha_s}{4\pi}\left( \nu^2
    e^{\gamma_E} \right)^{\!\epsilon}\; } \nonumber \\ &&
\quad \times  \left\{ \frac{d+\xi-3}{2}
    g(1,1) + \frac{1-\xi}{2} g(2,1)\right\} , \qquad
\\[1mm]
\lefteqn{
  \lambda_{B}^{(1)}(p^2,{p'}^2,q^2) = 4 N_c C_F \frac{\alpha_s}{4\,\pi} 
  \frac{d-(1-\xi)}{2} } \\
&& \times \! \left\{ \! q^2 j(1,\! 1,\! 1 ; p^2\!, p'^2\!, q^2)
                     + g(1,1) e^{\gamma_E\epsilon} \!
                     \left[ \! \left(\!\frac{\nu^2}{-p^2}\! \right)^{\!\!\epsilon}
                           \!\! + \!\! \left(\!\frac{\nu^2}{-{p'}^2}\!
                           \right)^{\!\!\epsilon} \right]\! \right\}\! , \nonumber
\\[2mm]
\lefteqn{
  \tilde \lambda_{B}^{(1)}(p^2,{p'}^2,q^2) = 2 N_c C_F \frac{\alpha_s}{4\,\pi} 
  }  \\
&& 
 \! \times \! \left\{
  \left[(d\!-\!2)^2 q^2\! - \!(d\!-\!2)(1\!-\!\xi) \left(p^2\!+\!{p'}^2\right) \right]
  \! j(1,\! 1,\! 1 ; p^2\!, p'^2\!, q^2)
\right.
 \nonumber \\
&& 
   -  2 \, (d\!- \!2) (1\!-\! \xi) \, g(1,1) e^{\gamma_E \epsilon}
               \left(\!\frac{\nu^2}{-{q}^2}\! \right)^{\!\!\epsilon}
  + \Big[ 2\, (1\!-\! \xi) g(1,2) 
\nonumber \\
&&
\left.
+ \! \left( (d\!-\!2)^2 \! + \! (d\!-\!4)(1\!-\!\xi) \right) \! g(1,1)\! \Big]
 e^{\gamma_E\epsilon} \!
                     \left[ \! \left(\!\frac{\nu^2}{-p^2}\!\!\right)^{\!\!\epsilon}
                           \!\! + \!\! \left(\!\frac{\nu^2}{-{p'}^2}\!\!
                           \right)^{\!\!\epsilon} \right]\! \right\}\! , \nonumber
\end{eqnarray}
where $\gamma_E$ is the
Euler-Mascheroni constant, $\nu$ the dimensional renormalization
scale and
\begin{equation}
   g(\nu_1, \nu_2) =
     \frac{\Gamma(\nu_1+\nu_2+\epsilon -\!2)
                            \Gamma(2-\epsilon-\nu_1) \Gamma(2-\epsilon-\nu_2)}
          {\Gamma(\nu_1) \Gamma(\nu_2) \Gamma(4-\nu_1-\nu_2-2 \epsilon)} .
\end{equation}
The function $j$ results from a massless triangle, via
\begin{eqnarray}
&&
 j(\nu_1, \nu_2 ,\nu_3; p_1^2, p_2^2, p_3^2) \equiv
 \left( \frac{i}{16 \pi^2} \right)^{\!\!-1} \!\!
  \left(\frac{\nu^2}{4\,\pi} e^{\gamma} \right)^{\!\!\epsilon} 
\nonumber \\
 &&
  \quad \times  \int \!\!\! \frac{d^d k}{(2\pi)^d} 
          \frac{1}{[-k^2]^{\nu_3} [-(k+p_1)^2]^{\nu_2}
            [-(k-p_2)^2]^{\nu_1}} , \qquad
\end{eqnarray}
with $p_3 = -(p_1+p_2)$.
Several cases have been evaluated in \cite{Usyukina:1994iw}
(our $j$ is essentially their $J$), in particular
\begin{eqnarray}
\lefteqn{
 j(1, 1, 1; p_1^2, p_2^2, p_3^2) =   } \\ \label{eq:j111}
&&      \left( \frac{\nu^2}{-p_3^2} e^{\gamma_E} \right)^{\!\epsilon} \;
     \frac{\Gamma(1+\epsilon)}{p_3^2} 
     \left( \Phi^{(1)}(x, y) + \epsilon \Psi^{(1)}(x, y)
          + \cO(\epsilon^2) \right) ,  \nonumber
\end{eqnarray}
where $x=p_1^2/p_3^2$ and $y=p_2^2/p_3^2$.
The functions $\Phi^{(1)}(x,y)$ and $\Psi^{(1)}(x,y)$ have been given in
\cite{Usyukina:1994iw} in terms of polylogarithms up to second and
third order, respectively.

\begin{figure}[t]
\centerline{
\includegraphics[width=.45\textwidth,viewport=40 130 720 620]{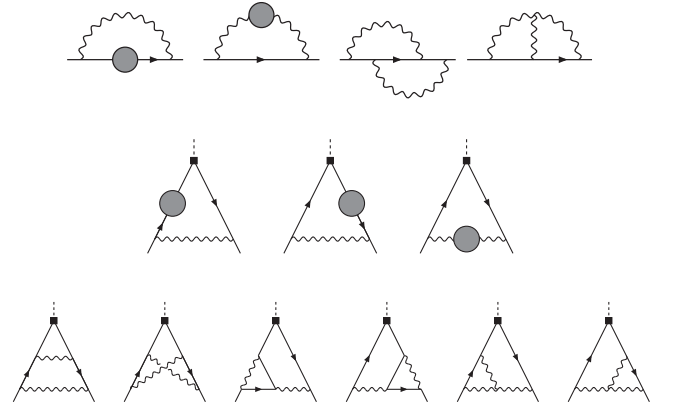}
}
\caption{Two-loop propagator and vertex diagrams. The grey blobs
indicate a sum over all one-loop corrections to a propagator,
the black boxes an insertion of a fermion bilinear}
\label{fig:vertdiag}
\end{figure}

\begin{figure}[t]
\centerline{
\includegraphics[width=.45\textwidth,viewport=50 440 700 610]{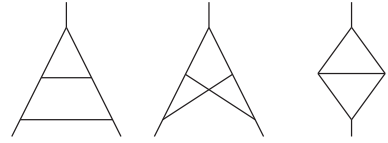}
}
\caption{Basic three- and two-point topologies: ladder,
  non-planar, propagator (from left to right)}
\label{fig:graphs}
\end{figure}

At the two-loop level, the relevant diagrams
are shown in Figure \ref{fig:vertdiag}.
They can be
represented in terms of three master ``topologies''
(Figure \ref{fig:graphs}), which may be called ``propagator'', ``ladder'',
and ``non-planar'',
with their propagators raised to general integer powers.
For the latter two topologies, irreducible numerators occur.
The set can be reduced by
standard reduction techniques and a systematic application
integration-by-parts (IBP) identities. For this we employ
the program FIRE  \cite{Smirnov:2008iw}, a public implementation of Laporta's
algorithm \cite{Laporta:2001dd} and the method of $S$-bases \cite{sbases}.
A subtle apsect of the IBP reduction is the occurence of
quadratic and simple poles in $\epsilon$ in the coefficients of the
resulting integrals.
In a two-loop computation, this leads to
poles of up to fourth order. On the other hand, the
Feynman diagrams have poles of at most second order, entirely of ultraviolet
origin. The spurious third- and fourth-order poles cancel, which
constitutes a check of the computation, but
they also imply a possible dependence
on terms up to ${\cal O}(\epsilon^3)$ in the $\epsilon$ expansion
of the master integrals remaining after the reduction.
In practice, we find that only known
master integrals \cite{Usyukina:1994iw,Usyukina:1994eg}
are needed, except for the unknown
${\cO}(\epsilon^2)$ part of $j(1,1,2+\epsilon)$. Denoting
\begin{eqnarray*}
 \lefteqn{ j(1, 1, 2+\epsilon; p_1^2, p_2^2, p_3^2) =
      \left( \frac{\nu^2}{-p_3^2} e^\gamma \right)^{\epsilon}
     \Gamma(1+\epsilon) (-p_3^2)^{-2-\epsilon} }
\\ && \times \frac{1}{2(1+\epsilon)x y}
       \left(\! - \frac{1}{\epsilon} + 2 \ln(x y)  
  \!\! + \epsilon \left[ \frac{\pi^2}{6}
              \! - \! 2\, (\ln^2 x + \ln^2 y) \right. \right.
\\ && \left. \left. - \ln x \ln y
              - \! 3 (1 \! - \! x \! - \! y) \Phi^{(1)}(x, y)\! \right]
             \! + \epsilon^2\, \Xi^{(1)}(x, y)
       + {\cal O}(\epsilon^3) \! \right) , \quad 
\end{eqnarray*}
we find that
\begin{eqnarray}
\lefteqn{  \Xi^{(1)}(x, x)\!  = \! \label{eq:betaxx}
      \frac{1}{2}\,\Omega^{(2)}(x, x)
      - \Omega^{(2)} \!\left(\!1, \frac{1}{x}\! \right)
            - (3\! -\! 6 x) \Psi^{(1)}(x, x)
 }
\nonumber \\ &&
+ \frac{11}{3} \ln^3 x  + 14\, \zeta(3)
      + \ln x \left(\! (3\! -\! 6 x) \Phi^{(1)}(x, x) - \frac{2}{3} \pi^2 
      \! \right), \quad\quad
\\
\lefteqn{  \Xi^{(1)}\!\left(\!1, \frac{1}{x} \right) \! = \! \label{eq:beta1x}
 - \frac{1}{2} \Omega^{(2)}(x, x)  \! + 3\, \Psi^{(1)}(x, x) }
\nonumber \\ && 
  -\frac{4}{3} \ln^3 x + 14 \,\zeta(3)
       + \ln x \left(\! \frac{\pi^2}{3} + \frac{9}{2} \Phi^{(1)}(x, x) \right)
       . \qquad
\end{eqnarray}
The function  $\Omega^{(2)}$ arises in evaluating ladder master integrals
\cite{Usyukina:1994iw,Usyukina:1994eg} and is given there in terms of
polylogarithms.
Combining all terms
and $\msbar$-renormalizing the gauge coupling and gauge parameter,
we obtain
\begin{widetext}
\begin{eqnarray}  \label{eq:Cmresult}
\lefteqn{ C_m^{\rm RI/SMOM}(\omega)=
 \; 1 + \frac{\alpha_s}{4\pi} \, C_F \left(\frac{3+\xi}{2}
         \, \Phi^{(1)}\!\left(\frac{1}{\omega},
           \frac{1}{\omega}\right) -4 - \xi  + 3\,\ln r \! \right) 
%
%
 + \left(\frac{\alpha_s}{4\pi} \right)^{\!\!2} \, C_F \,
\left\{  N_c \left( - \frac{2513}{48} - \frac{3 \, \xi}{2} -
  \frac{\xi^2}{4}  + 12 \, \zeta(3)  \right. \right. }
\nonumber \\ && \quad 
\left.
 + \frac{307 + 6\,\xi^2}{12} \ln r
 - \frac{13}{4} \ln^2 r
   + \left[ \frac{301}{24}\! +\! \frac{3\,\xi}{4}\! -\! \frac{\xi^2}{8}
   \!- \! \frac{13+\xi^2}{4} \ln r
   - \frac{7+3\,\xi}{4} \ln \omega \right]
        \Phi^{(1)}\!\left(\frac{1}{\omega}, \frac{1}{\omega}\right)
  + \frac{9+6\,\xi+\xi^2}{8}
       \Phi^{(1)}\!\left(\frac{1}{\omega}, \frac{1}{\omega}\right)^{\!2} \!\!
 \right.
\nonumber \\ && \left.
\quad +\, \omega \; \Phi^{(2)} \! \left(1, \omega \right)
   - \frac{3 + \xi}{2} \, \Phi^{(2)}\!\left(\frac{1}{\omega}, \frac{1}{\omega}\right) \right)
%
%
 \; + \; n_f \left( \frac{83}{12}
 + \! \left[ \ln r - \frac{5}{3} \right]
         \Phi^{(1)}\!\left(\frac{1}{\omega}, \frac{1}{\omega}\right)
 - \frac{13}{3} \ln r + \ln^2 r 
          \! \right)
\nonumber \\ && \quad  \nonumber
      + \frac{1}{N_c} \left( - \frac{19}{16} -2 \, \xi -
      \frac{\xi^2}{2}
      + \left[ \frac{7}{2} + \xi + \frac{\xi^2}{2}
      - \frac{9+3\,\xi}{4} \ln r
      + \frac{5+3\,\xi}{4} \ln \omega \right]
            \Phi^{(1)}\!\left(\frac{1}{\omega}, \frac{1}{\omega}\right)
      + \frac{21+6\,\xi}{4} \ln r
 - \frac{9}{4} \ln^2 r
\right. \\
&&  \quad \left.  \left.
 + \frac{1+\xi}{2}\, \Phi^{(2)}\!\left(\frac{1}{\omega}, \frac{1}{\omega}\right) + \frac{1}{2}\, \Omega^{(2)}\!\left(\frac{1}{\omega}, \frac{1}{\omega}\right) -
   \Omega^{(2)} (1, \omega )
 - \left[ \frac{5}{8} +\frac{3\,\xi}{4}+\frac{\xi^2}{8}
             + \frac{1}{\omega} \right]
       \Phi^{(1)}\!\left(\frac{1}{\omega}, \frac{1}{\omega}\right)^{\!2}
 \right) \right\} + {\cal O} \left( \alpha_s^3 \right),
\end{eqnarray}
\begin{eqnarray}  \label{eq:Cqgamresult}
 \lefteqn{ C_q^{\smomgam}(\omega)=
 \; 1 + \frac{\alpha_s}{4\pi} \, C_F \left(1-\frac{3\,\xi}{2} + \xi \ln r
      - \frac{1\!-\!\xi}{2} \ln \omega + \frac{\omega\!-\!1\!+\!\xi}{2\,\omega} 
         \, \Phi^{(1)}\!\left(\frac{1}{\omega},
           \frac{1}{\omega}\right) \right) 
%
%
 }
\nonumber \\
&&  + \left(\frac{\alpha_s}{4\pi} \right)^{\!\!2} C_F 
\left\{ 
N_c \left(  \!  - \frac{71}{144} - \frac{35\,\xi}{4} 
  - \frac{5\,\xi^2}{8}
  - \frac{3-9\,\xi}{2} \zeta(3)
  + \left[ \frac{11}{6} + \frac{19\,\xi}{4} + \frac{\xi^2}{4}
         + \left(\frac{11}{6} - \xi\right) \ln \omega  \right] \ln \, r
  - \frac{3\,\xi}{4}\ln^2\,r
\right. \right.
\nonumber \\
&& \left. \left. \quad
+ \left[ \frac{223\,\omega-259}{72\,\omega}
          + \frac{\omega+20}{8\,\omega}\xi
          - \frac{\xi^2}{8\,\omega}
          + \frac{22(1-\omega) + (3\,\omega - 12)\xi}{12\,\omega} \ln  r
          + \frac{1 + (\omega-2)\xi + \xi^2}{4\,\omega} \ln \omega
\right] \Phi^{(1)}\!\left(\frac{1}{\omega}, \frac{1}{\omega}\right)
\right. \right.
\nonumber \\
&& \left. \left. \quad
- \frac{259 - 180\,\xi + 9\,\xi^2}{72}\ln\omega
+ \frac{(1-\xi)^2}{8} \ln^2 \omega
+ \frac{(1 - \omega - \xi)^2}{8\,\omega^2}
          \Phi^{(1)}\!\left(\frac{1}{\omega}, \frac{1}{\omega}\right)^2
+ \frac{1-\xi}{4}\, \Omega^{(2)}(1, \omega)
- \frac{3-\xi}{8}\, \Omega^{(2)}\!\left(\frac{1}{\omega},\frac{1}{\omega}\right)
\right. \right.
\nonumber \\
&& \left. \left. \quad
+ \frac{\omega}{2} \Phi^{(2)}(1, \omega)
+ \frac{3 - 2 \omega - \xi}{2\,\omega}
          \Phi^{(2)}\!\left(\frac{1}{\omega}, \frac{1}{\omega}\right)
\right) \;\; + n_f \left( \frac{5}{36} - \frac{1+\ln\omega}{3}\ln r +
  \frac{5}{9}\ln\omega + \frac{(1-\omega) (5 -3\ln\,r)}{9\,\omega}
     \Phi^{(1)}\!\left(\frac{1}{\omega}, \frac{1}{\omega}\right)
\right) \right.
\nonumber \\
&& \left. \quad
+ \frac{1}{N_c} \left(
  \frac{1}{16} + \frac{\xi}{2}
  - \frac{7\xi^2}{8}
  + 3\, (1-\xi)\, \zeta(3)
  - \frac{\xi^2}{4} \ln^2 r
  + \frac{1-3\,\xi+2\,\xi^2}{4} \ln\omega
  + \left[
  \frac{3-2\,\xi+3\,\xi^2 }{4}
            + \frac{\xi(1\!-\!\xi)}{4} \ln \omega \right] \ln r
\right. \right.
\nonumber \\
&& \left. \left. \quad
  - \frac{(1-\xi)^2}{8}\ln^2\omega
  + \left[ \frac{13}{8} + \frac{1}{4\,\omega} -
    \frac{6+\omega}{8\,\omega}\xi + \frac{\xi^2}{2\,\omega}
    + \frac{\xi(1-\omega-\xi)}{4\,\omega} \ln r 
    - \frac{1+\omega + \xi (\omega-2) + \xi^2}{4\,\omega} \ln\omega
\right] \Phi^{(1)}\!\left(\frac{1}{\omega},
           \frac{1}{\omega}\right) 
\right. \right.
\nonumber \\
&& \left. \left. \quad
 - \frac{1 + \omega\,(2 - \omega) - 2\,\xi(1-\omega) + \xi^2}{8\,\omega^2} 
           \Phi^{(1)}\!\left(\frac{1}{\omega}, \frac{1}{\omega}\right)^{\!2}   
 - \frac{1-\xi}{2\,\omega}
        \Phi^{(2)}\!\left(\frac{1}{\omega}, \frac{1}{\omega}\right)
  - \frac{1-\xi}{2} \, \Omega^{(2)} \! \left(1, \omega \right)
 \right)
 \right\} + {\cal O}(\alpha_s^3) ,
\end{eqnarray}
\begin{eqnarray}  \label{eq:Cmgamresult}
 \lefteqn{ C_m^{\smomgam}(\omega, \omega)=
 \; 1 + \frac{\alpha_s}{4\pi} \, C_F \left(-5 - \frac{\xi}{2} + 3 \ln r
      + \frac{1-\xi}{2} \ln \omega
       + \frac{1+2\omega + (\omega-1)\xi}{2\,\omega} 
         \, \Phi^{(1)}\!\left(\frac{1}{\omega}, \frac{1}{\omega}\right) \right)  } \nonumber \\
&&
%
%
 + \left(\frac{\alpha_s}{4\pi} \right)^{\!\!2} \, C_F \,
\left\{
N_c \left(
- \frac{8539}{144}-\frac{3\xi}{4} - \frac{\xi^2}{8}
 + \frac{33-3\xi}{2} \zeta(3)
+ \frac{151 - 54 \xi - 9 \xi^2}{72} \ln\omega
+ \left[ \frac{111 + \xi^2}{4}
         + \frac{ 3\xi^2 - 13}{12} \ln\omega \right] \ln r
\right. \right.
\nonumber \\
&& \left. \left. \quad
 - \frac{13}{4}\ln^2 r
+ \left[ \frac{151 + 734 \omega}{72\,\omega} + \frac{2\omega-3}{4\,\omega} \xi
    - \frac{\xi^2}{8\,\omega}
    - \frac{13+8\xi+\xi^2}{8} \ln\omega
    + \frac{3 (1-\omega) \xi^2 - 13 (1 + 2 \omega)}{12\,\omega} \ln r
\right] \Phi^{(1)}\!\left(\frac{1}{\omega}, \frac{1}{\omega}\right)
\right. \right.
\nonumber \\
&& \left. \left. \quad
+ \frac{3+6\,\omega + (5\,\omega-2)\xi + (\omega-1) \xi^2}{8\,\omega}
     \Phi^{(1)}\!\left(\frac{1}{\omega}, \frac{1}{\omega}\right)^2
+ \frac{\omega}{2} \Phi^{(2)}(1, \omega )
- \frac{3 + \omega + (\omega - 1) \xi}{2\,\omega}
      \Phi^{(2)}\!\left(\frac{1}{\omega}, \frac{1}{\omega}\right)
- \frac{1-\xi}{4} \Omega^{(2)}(1, \omega)
\right. \right.
\nonumber \\
&& \left. \left. \quad
+ \frac{3-\xi}{8} \Omega^{(2)}\!\left(\frac{1}{\omega}, \frac{1}{\omega}\right)
\right)
+ n_f \left( \frac{307}{36} + \ln^2 r - \frac{5}{9}\ln\omega
+ \frac{\ln\omega-15}{3}\ln r
+ \frac{(1+2\omega)\, (3\ln r - 5)}{9\omega}\,
       \Phi^{(1)}\!\left(\frac{1}{\omega}, \frac{1}{\omega}\right)
\right)
\right.
\nonumber \\
&& \left.
+ \frac{1}{N_c} \left(
- \frac{65}{16} -\xi - \frac{\xi^2}{4} - 3 (1-\xi)\, \zeta(3)
- \frac{9}{4} \ln^2 r
+ \frac{5-4\xi-\xi^2}{4}\ln\omega
+ \left[\frac{27}{4} +\frac{3\xi}{4} -\frac{3(1-\xi)}{4} \ln\omega \right] \ln r
- \frac{1+\xi}{2}\, \Omega^{(2)} \left(1, \omega \right)
\right. \right.
\nonumber \\
&& \left. \left. \quad
+ \frac{1}{2}\, \Omega^{(2)}\!\left(\frac{1}{\omega}, \frac{1}{\omega}\right)
+ \frac{1-\xi+\omega (1+\xi)}{2\omega} \,
      \Phi^{(2)}\!\left(\frac{1}{\omega}, \frac{1}{\omega}\right)
+ \left[ \frac{-7+2\xi+\xi^2}{8\omega} - \frac{4+5\xi+\xi^2}{8}
\right] \Phi^{(1)}\!\left(\frac{1}{\omega}, \frac{1}{\omega}\right)^2
\right. \right.
\nonumber \\
&& \left. \left. \quad
+ \left[ \frac{9+10\xi+3\xi^2}{8} + \frac{5-4\xi-\xi^2}{4\,\omega}
+ \frac{3(1-\omega) \xi - 3 - 6\omega}{4\,\omega}\ln r
+ \frac{11+8\xi+\xi^2}{8} \ln\omega
\right] \Phi^{(1)}\!\left(\frac{1}{\omega}, \frac{1}{\omega}\right)
\right)
 \right\} + {\cal O}(\alpha_s^3) ,
\end{eqnarray}
\end{widetext}
where $r = \mu^2/\nu^2$, $n_f$ is the number of quark
flavours, and $\Phi^{(2)}$
is given in terms of polylogarithms
in \cite{Usyukina:1994iw}.
The function
$\Psi^{(1)}$ has dropped out of the final results.
We do not know the
origin of this cancellation, involving many
different terms, including the $\cO(\epsilon)$ one-loop terms.
As an elk test,  setting $\mu=\nu$ in \eq{eq:Cmresult}
and taking $\omega\! \to\! 0$,
we recover $C_m^{\ripmom}\!$
\cite{Chetyrkin:1999pq,Gracey:2003yr}.
The $\cO(\alpha_s)$ terms in \eq{eq:Cmresult}--\eq{eq:Cmgamresult}
agree with \cite{Sturm:2009kb} (for $\omega=1$ and $r=1$).
The most general form $C_m^\smomgam(\omega, \omega')$, defined in
\eq{eq:CmSMOMgam}, can be obtained from \eq{eq:Cqgamresult} and
\eq{eq:Cmgamresult} as
\begin{equation}
  C_m^\smomgam\!(\omega, \omega') \!=\!
     \frac{C_q^\smomgam\!(\omega)}{C_q^\smomgam\!(\omega')}
     C_m^\smomgam\!(\omega,\omega).
\end{equation}

The mass and field anomalous dimensions in the two schemes are
easily obtained by substituting the expressions \eq{eq:Cmresult},
\eq{eq:Cqgamresult}, and \eq{eq:Cmgamresult}, as well
as $C_q^{\rm SMOM} = C_q^\ripmom$ \cite{Chetyrkin:1999pq,Gracey:2003yr}
and the well-known two-loop
$\beta$-function into \eq{eq:admm} and \eq{eq:admq}.
More explicitly, denoting
\begin{eqnarray}
  \beta(\alpha_s) &=&
    - \beta^{(0)} \left( \frac{\alpha_s}{\pi} \right)^2
    - \beta^{(1)} \left( \frac{\alpha_s}{\pi} \right)^3 + {\cal O}(\alpha_s^4) ,
\\
  C^X_p &=& 1 + C^{X(1)}_p \left( \frac{\alpha_s}{4\,\pi} \right)
        + C^{X(2)}_p \left( \frac{\alpha_s}{4\,\pi} \right)^2
        + {\cal O}(\alpha_s^3) , \qquad
\end{eqnarray}
where $p=m$ or $p=q$, $X=$ RI/SMOM or $\smomgam$,
$\beta^{(0)} = (11 N_c - 2 \, n_f)/12$,
$\beta^{(1)} = (34\, N_c^2 - 10\, N_c n_f - 6\, C_F n_f)/48 $, and
the remaining coefficients can be read off
\eq{eq:Cmresult}--\eq{eq:Cmgamresult}, we have to NNLO:
\begin{eqnarray}  \label{eq:admconv}
  \gamma^X_p &=& \gamma^\msbar_p
      + 4 \, \beta^{(0)} C^{X(1)}_p \left( \frac{\alpha_s}{4\,\pi} \right)^2
\nonumber \\
&&      + 4 \left\{ \! \beta^{(0)} \! \left[2 \, C^{X(2)}_p
                  \!-\! \big(C^{X(1)}_p\big)^2 \right] \!
                  + \! 4\, \beta^{(1)} C^{X(1)}_p \! \right\} \!
           \left( \frac{\alpha_s}{4\,\pi} \right)^{\!3}
\nonumber \\[1mm]
&&      + \Delta_p^X  + {\cal O}(\alpha_s^4) .
\end{eqnarray}
Here
\begin{equation}
  \Delta^X_p \equiv \delta(\alpha_s, \xi) \, [C^X_p]^{-1}\frac{\partial C^X_p}{\partial \xi},
\end{equation}
which vanishes in the Landau gauge,
is again straightforward to evaluate to ${\cal O}(\alpha_s^3)$ from
\eq{eq:Cmresult}--\eq{eq:Cmgamresult} and the perturbation expansion of
$\delta$ defined in \eq{eq:admdef}.
Gauge invariance implies $\delta = \gamma_A \xi$, where
$\gamma_A$ is the anomalous dimension of the gluon field (defined
analogously to \eq{eq:admdef})
\cite{Tarasov:1976ef,Larin:1993tp,Chetyrkin:2004mf}, giving
\begin{eqnarray}
\delta &=& \frac{\alpha_s}{4 \pi}\, \xi\! \left(
\frac{13\! -\! 3\, \xi}{6}  N_c - 
\frac{2}{3} n_f \!
\right)
\\
&& \!\!\! + \left(\frac{\alpha_s}{4 \pi}\right)^{\!2} \! \xi
\! \left(
\frac{59\! - \!11\,\xi\!-\!2\,\xi^2}{8} N_c^2 \!
-\!\frac{7 N_c^2 \!-\! 2 }{2\,N_c}\,  n_f
\!\!\right) \! + {\cal O}(\alpha_s^3) . \nonumber
\end{eqnarray}

\section{Phenomenology}
\begin{figure}[t]
\centerline{
\includegraphics[width=0.45\textwidth]{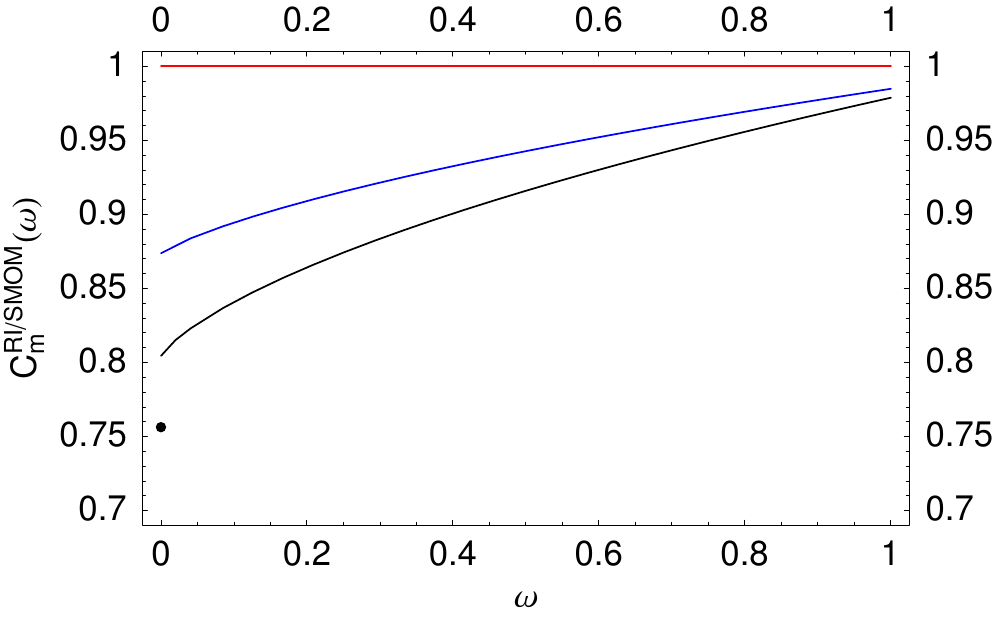}
}
\caption{Conversion factors $C_m^{\rm RI/SMOM}$ as
function of $\omega=q^2/p^2$
at LO (top/red), NLO (middle/blue), and NNLO (bottom/black),
and $C_m^\ripmom=C_m^{\rm RI/SMOM}(0)$ at NNNLO (dot).}
\label{fig:plotCms}
\end{figure}
\begin{figure}[t]
\centerline{
\includegraphics[width=0.45\textwidth]{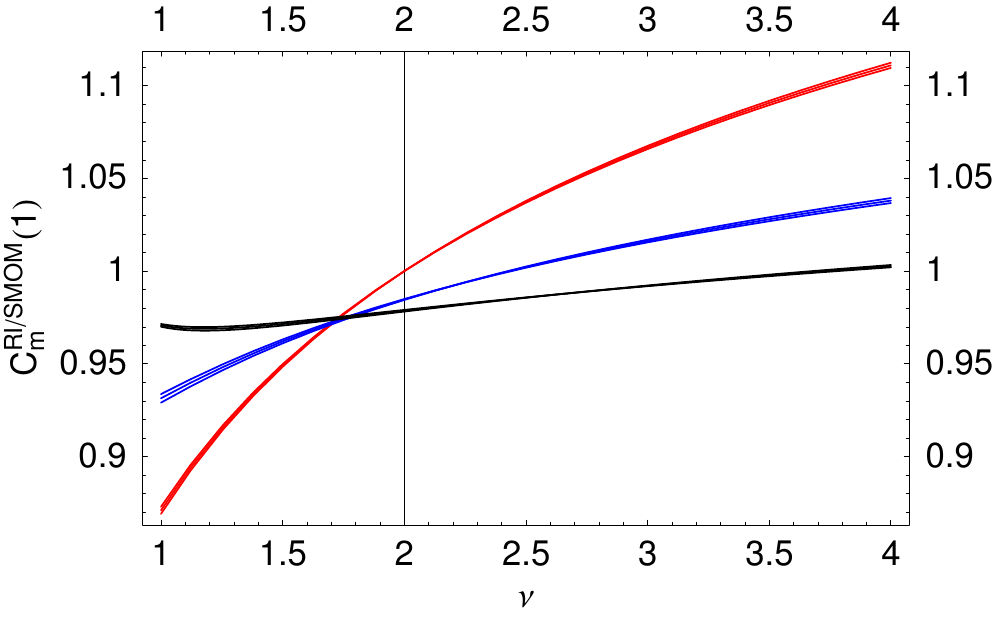}
}
\caption{Residual matching-scale dependence of the
conversion factor $C_m^{\rm RI/SMOM}$ at $\omega=1$ at
LO (red), NLO (blue), and NNLO (black).
}
\label{fig:plotCRISMOM}
\end{figure}
To explore the phenomenological consequences of our result
for QCD with three dynamical light quarks (as in nature, and in modern
unquenched simulations), we set $n_f=3$.
Figure \ref{fig:plotCms} shows the
conversion factor $C_m(\omega)$ in the Landau gauge.
We observe that the NNLO correction, like the NLO term,
is very small at the SMOM point $\omega=1$. This is in contrast
to the $\ripmom$ scheme $\omega=0$, where even the
next-to-next-to-next-to-leading-order (NNNLO)
correction \cite{Chetyrkin:1999pq,Gracey:2003yr}
is large (dot in the Figure).
To estimate the effects from uncomputed $\cO(\alpha_s^3)$ terms, we vary
the renormalization scale (matching scale) $\nu$ used in the
conversion and evolve $C_m^{\rm RI/SMOM}(\omega=1; \nu)$ to the fixed scale
$\mu=2$ GeV, which gives a
formally $\nu$-independent number \cite{rundec,Chetyrkin:2000yt}.
The result is shown in Figure \ref{fig:plotCRISMOM}.
The width of each band, due to the uncertainty on
$\alpha_s(M_Z)=0.1184 \pm 0.0007$ \cite{Bethke:2009jm}, is almost
negligible. This
is a consequence of the smallness of the NLO and NNLO corrections.
We observe that the NNLO result is almost
scale-independent.
Alternatively, we can convert the $\msbar$ mass to
the RGI quark mass employing the relevant expressions
in \cite{Chetyrkin:2000yt},
which is also scale-independent. The result is
similarly stable under scale variation, but the $\alpha_s$ dependence
is a bit more pronounced. A slightly larger residual scale dependence
is found for the $\smomgam$ scheme.
Numerically, we obtain
\begin{eqnarray} \label{eq:qmass} \nonumber
&&  m^\msbar(2\,\GeV) 
\\  \nonumber
&&\qquad \; \;\, = \left(0.979^{+0.024}_{-0.010}\big|_{\rm h.o.}{}^{+0.001}_{-0.001}\big|_{\alpha_s}\right)
     m^{\rm RI/SMOM}(2\,\GeV) , \quad \\[1mm]    \nonumber
&&\qquad \; \;\, =
  \left(0.932^{+0.030}_{-0.021}\big|_{\rm h.o.}{}^{+0.003}_{-0.003}\big|_{\alpha_s}\right)
     m^{\smomgam}(2\,\GeV) , \quad \\[1mm]
&& m^{\rm RGI} =
    \left(2.53^{+0.05}_{-0.02}\big|_{\rm h.o.}{}^{+0.02}_{-0.02}\big|_{\alpha_s}\right)
       m^{\rm RI/SMOM}(2\,\GeV)  \nonumber \\
&& \qquad \; \;\, =
    \left(2.41^{+0.07}_{-0.04}\big|_{\rm h.o.}{}^{+0.03}_{-0.03}\big|_{\alpha_s}\right)
       m^{\smomgam}(2\,\GeV) ,
\nonumber 
\end{eqnarray}
corresponding to a perturbative uncertainty of less than 2\%,
or about 2 MeV for the strange quark mass, when converting from
the RI/SMOM scheme,
and about 3\% for the $\smomgam$ scheme. As also the absolute size
of the NLO and NNLO corrections is larger for the $\smomgam$ scheme,
we advocate the use of the RI/SMOM scheme together with an appropriate
error estimate in extracting results for the light quark masses.

\section{Conclusion}
We have computed the RI/SMOM $\to \msbar$
and $\smomgam$ $\to \msbar$ conversion factors
for the quark mass to NNLO and shown that the RI/SMOM and $\smomgam$ schemes,
designed to reduce sensitivity to low-energy non-perturbative physics,
are perturbatively very well behaved, too. These schemes thus may be used
to extract quark masses with percent-level accuracy from numerical
lattice QCD. An important question is whether the same holds true
for other quantities of interest, such as $B_K$ and other hadronic
matrix elements.

\begin{acknowledgments}
We are happy to thank Chris Sachrajda for a talk and conversations
raising our interest in the topic and him and Andrzej Buras for
comments on the manuscript. M.\ G.\ thanks
the Galileo Galilei institute for hospitality and the INFN for
support during a stay.
\end{acknowledgments}

\paragraph{Note:}
After the initial submission of this manuscript to the Arxiv,
Ref.\ \cite{Almeida:2010ns}
appeared, whose authors compute the conversion factor
$C_m$ at NNLO for the symmetric renormalization point $\omega=1$, where they
confirm our result. They also give the corresponding field and mass
conversion factors $C_q^{\smomgam}$ and $C_m^{\smomgam}$ for the
$\smomgam$ scheme, as well as expressions for the
NNLO anomalous dimensions in both schemes. In this revised version,
we have given expressions for those quantities, as well.
Specialising to $\omega=\omega'=r=1$, our results for the $C^X_q$ 
agree with the results in \cite{Almeida:2010ns}. Setting further $\xi=0$, we
agree with the results for $\gamma_q^X$ given there, up to a
global sign difference \cite{admsign}.

\end{document}